\documentclass[%
 reprint,
 amsmath,amssymb,
 aps,
pra,
floatfix,
]{revtex4-2}

\usepackage{graphicx}
\usepackage{enumerate}
\usepackage{color}

\begin{document}


\title{Scattering Times of Quantum Particles from the Gravitational Potential, and Equivalence Principle Violation}

\author{Durmu{\c s} Demir}
 \affiliation{Sabanc{\i} University, Faculty of Engineering and Natural Sciences, 34956 Tuzla, {\.I}stanbul, Turkey}

\date{\today}

\begin{abstract}
Universality of motion under gravity, the equivalence principle, is violated for quantum particles. Here, we study time it takes for a quantum particle to scatter from the gravitational potential, and show that the scattering time, formulated here using the opportune Bohmian formulation, acts as an indicator of the equivalence principle violation. The scattering times of wavepackets are distinctive enough to distinguish between the Bohmian and Copenhagen interpretations. The scattering time of mono-energetic stationary states, formulated here as a modification of the Bohmian time by probability undercurrents, turns out to be a sensitive probe of the equivalence principle violation.  We derive the quantum scattering times, and analyze equivalence principle violating terms systematically. We discuss the experimental setup needed for measuring the violation, and describe implications of a possible measurement for time in quantum theory, including the tunneling time.
\end{abstract}

\maketitle

\section{Introduction}
Equivalence of the inertial mass $m_I$ and the gravitational charge $m_G$ of each and every particle \cite{free-fall-th1,free-fall-th2} renders the Newtonian dynamics purely geometrical. This equivalence has been tested for centuries, and has now reached some $10^{-15}$ relative accuracy \cite{free-fall-exp1,free-fall-exp2}.  It implies that all  small bodies (classical particles) fall at the same time if released from the same height with the same velocity. This universality is a fundamental aspect of  Newtonian dynamics under gravity, and provides a universal time scale testable  with distinct bodies \cite{free-fall-exp1}. 

The universality above is expected  to be invalid for quantum particles since their masses continue to appear in the Schroedinger equation even when $m_I\equiv m_G$ \cite{sciama,wadati,wadati2,wadati3,emelyanov,hu}. The thing is that $\hbar$ comes out of the woodwork as a new constant with the dimension of mass-distance-velocity and, in consequence, quantum dynamics remain non-geometrical with or without the equivalence $m_I\equiv m_G$. The free-fall time of quantum particles have been analyzed by Davies \cite{davies} by utilizing the Peres clock \cite{slacker,peres,peres2}, by Viola and Onofrio \cite{onofrio}  by using the semiclassical dynamics,  by Ali and others \cite{alis}  by considering wavepackets, by Flores and Galapon \cite{operator} by making use of a time operator \cite{operator2}, and by Seveso and others by utilizing the information-theoretic methods \cite{seveso1,seveso2}. These studies use different methods but agree on the existence of a finite deviation from the universal free-fall time of the classical dynamics. As will be analyzed in detail in the sequel, deviation from universal free-fall time or, equivalently, violation of the equivalence principle is characterized by the ratio $\hbar/m$. This quantity  disappears and gives then way to universality  only in the classical limit namely only when action of the particle is significantly sizable than $\hbar$. 

The crux of the problem is that in quantum theory time is not an observable represented by a hermitian operator, and there is thus no baseline methodology to calculate temporal intervals \cite{challinor,abolhasani,busch,field}. The resolution, according to various proposals \cite{hilgevoord,muga}, is that the march of  time in quantum systems should be defined in terms of the changes in certain representative observables  (position, momentum, spin orientation and the like \cite{hilgevoord,busch}). The values of these observables act as duration markers \cite{field} but choice of the observables depends on how the particle is modeled or perceived (like, for example, wavepackets or energy eigenstates). In general, however, an unambiguous definition and elucidation of time is necessary for both the foundations and applications of the quantum theory \cite{slacker,calcada,field}. 

In search for a proper reification of time, the Copenhagen (standard) and Bohmian \cite{bohm} interpretations provide two alternative routes \cite{field}. In the Copenhagen interpretation, quantum particles do not possess well-defined trajectories. In the de Broglie-Bohm interpretation, on the other hand, quantum particles possess well-defined trajectories controlled by their probability flows \cite{bohm,bohmian,sonego}. The two interpretations are empirically equivalent in that they have different views on reality and yet they give identical results on main physical questions \cite{bohmian}. This equivalence of theirs gets, however, disrupted once trajectories and probability flows of particles are concerned. In this regard, one phenomenon on which the two interpretations disagree is the probability backflow (having momentum and probability current in opposite directions) which plagues the standard interpretation \cite{backflow1} but  does not occur at all in the Bohmian interpretation. In this sense, future experiments on probability backflow \cite{backflow2,backflow3,bohmian2} may differentiate between the two interpretations. 

Another occasion in which the two interpretations differ concerns the notion of the quantum travel time.  In the standard interpretation (where particles evolve as distributions until they collapse indeterministically under some measurement process) the march of time has been defined variously  (see the reviews \cite{hilgevoord,muga,phase-2}) by using different observables as markers. In the Bohmian interpretation (where particles follow a well-defined trajectory under the control of the Schroedinger equation) the march of time can be expressed uniquely in terms of their positions \cite{bohm,sonego,bohmian}. (Comprehensive explorations in \cite{field,cushing-1,cushing-2} shed light on different aspects of the two interpretations in regard to the problem of time in quantum theory.)
 
In this work, we shall study time it takes for a quantum particle to scatter from its own gravitational potential, and use this scattering time to determine/measure the  equivalence principle violation. In the setup we consider, quantum particles are shot upwards and their return times are recorded such that deviations of the recorded times from the classical universal flight time will be an indicator of the equivalence principle violation. We shall study this problem within the de Broglie-Bohm interpretation of the quantum theory \cite{bohm,bohmian,sonego,field}. The scattering times we will compute will be average times in that the Bohmian time formula  involve integrations over probability and probability current densities. 

In Sec. II below, we study scattering times of wavepackets (having classical analogs), and show that dispersion of the wavepacket is the main source equivalence principle violation. We compare our finding with the Copenhagen result (with time operator method \cite{operator}), and conclude that future experiments may be able to probe what interpretation of quantum theory is realized in nature.  

 In Sec. III, we study flight times of mono-energetic stationary-state particles (having no classical analogs). We extend the Bohmian time formula of Sec. II to probability undercurrents to obtain a Bohmian-inspired time formula. We show that the Bohmian-inspired time formula is tailor-made for such states. Therein, we determine  scattering times of quantum particles in terms of the corresponding classical scattering times. We apply the Bohmian-inspired time formula in both of the classically-allowed and classically-forbidden regions, study its short- and high-flight limits, and reveal the sources of equivalence principle violation. We find that quantum particles spend time behind the classical turning point during their penetration into and withdrawal from the gravitational potential barrier. 
 
 In Sec. IV, we discuss state-of-the-art experimental situation, and discuss how the quantum scattering times (for both the wavepackets and stationary states) can be tested in cold atom experiments. We also discuss their  implications for applications and foundations of  quantum mechanics. 
 
 In Sec. V we conclude.

\section{Quantum Scattering Time: Wavepackets} \label{secwaveacket}
In vacuum (negligible friction), small bodies (macroscopic particles with negligible tidal forces) obey Newton's motion equation 
\begin{eqnarray}
 \frac{d^2 z(t)}{dt^2}=-g
 \label{newton}
\end{eqnarray}
for a uniform gravitational field $g$ pointing in negative $z$ direction. This equation is universal (same for all particles) thanks to the equality $m_I = m_G\equiv m$ between the inertial mass $m_I$ and the gravitational charge $m_G$. As a result, all classical particles,  tossed upwards from a vertical position $z=z_i$ with initial velocity $v_i$, follow one and the same trajectory  
\begin{eqnarray}
z_c(t)=z_i +v_i t - \frac{1}{2} g t^2
\label{z-class}
\end{eqnarray}
as a solution of (\ref{newton}). This means that rise of the particle comes to a halt at the moment $t_\cap=v_i/g$ corresponding to a height of $z_c(t_\cap)\equiv z_\cap=z_i + v_i^2/2g$. This height $z_\cap$ is the turning point. 

Universality of the classical motion above is not expected to  hold for quantum particles. The reason is that, unlike the Newtonian motion equation (\ref{newton}), the Schroedinger equation 
\begin{eqnarray}
-\frac{\hbar^2}{2 m} \frac{\partial^2 \Psi(t,z)}{\partial z^2} + V(z)\Psi(t,z) = i\hbar \frac{\partial \Psi(t,z)}{\partial t}
\label{sch-time}
\end{eqnarray}
depends explicitly on the particle masses irrespective of if $m_I\equiv m_G$ or $m_I\not\equiv m_G$. This non-universality has the meaning that the equivalence principle is violated for quantum particles.  In fact, with the gravitational potential energy 
\begin{eqnarray}
V(z)= m g z
\label{pot}
\end{eqnarray}
the Schroedinger equation (\ref{sch-time}) is seen to invariably involve the dimensionful parameter $\hbar/m$. It admits different solutions, one of which being the wavepacket solution  \cite{wadati,wadati2,wadati3}
\begin{eqnarray}
\Psi(t,z)&=&\left(\frac{d}{\sqrt{\pi} D^2}\right)^{1/2} \exp \left\{-\frac{(z-z_c)^2}{2D^2}+\frac{m}{i\hbar} z_i v_i\right\}\nonumber\\
&\times& \exp \left\{-\frac{m}{i\hbar}\left(z-\frac{v_i t}{2}\right)\left(v_i - gt\right) + \frac{m g^2}{i\hbar} t^3\right\}
\label{wpacket}
\end{eqnarray}
characterized by the time-varying width $D^2=d^2 + \frac{i\hbar t}{m}$. It is obviously a non-stationary-state  as its phase is not linear in time $t$. It is an approximation to the notion of particle, and possesses  classical analog in that its center $z=z_c$ follows the Newtonian motion equation (\ref{newton}). 

The foremost feature of the Bohmian mechanics is that it ascribes trajectories $z(t)$ to quantum particles such that
\begin{eqnarray}
\frac{d z}{dt}= \frac{J(t,z)}{R(t,z)}
\label{eom-bohm}
\end{eqnarray}
under the control of the Schroedinger equation (\ref{sch-time}). In this regard,  $R(t,z)=\Psi^\ast(t,z)\Psi(t,z)$ is the probability density, and 
\begin{eqnarray}
\!\!\!\!\!\!\!J(t,z)=\frac{\hbar}{2 m i}\! \left(\!\Psi^\ast(t,z)  \frac{d}{d z}\Psi(t,z)-\Psi(t,z)  \frac{d}{d z}\Psi^\ast(t,z)\!\right)
\label{prob-current}
\end{eqnarray}
is the probability current density. They satisfy the  continuity equation 
\begin{eqnarray}
\frac{\partial R(t,z)}{\partial t} + \frac{\partial J(t,z)}{\partial z} =0
\label{continuity}
\end{eqnarray}
which ensures the conservation of probability. Their ratio 
\begin{eqnarray}
\frac{J(t,z)}{R(t,z)} = \frac{t \left(z-z_c(t)\right)}{\frac{m^2 d^4}{\hbar^2} +t^2}  + v_i-gt
\label{eom-bohm-wpacket}
\end{eqnarray}
reveals the non-classical features of  wavepacket (\ref{wpacket}). In fact, with this current-to-probability ratio  the Bohmian equation (\ref{eom-bohm}) takes the compact form
\begin{eqnarray}
\frac{d}{d t} \left(\frac{z-z_c(t)}{\left(1+\frac{\hbar^2 t^2}{m^2d^4}\right)^{\frac{1}{2}}}\right) = 0
\label{bohmian-wpacket-eqn}
\end{eqnarray}
showing explicitly that deviation from the classical solution $z=z_c(t)$ occurs due solely to the wavepacket dispersion ($\hbar t/m d^2$). In fact, this equation acquires the solution
\begin{eqnarray}
z(t)=z_c(t) + \ell \left(1+\frac{\hbar^2 t^2}{m^2d^4}\right)^{\frac{1}{2}}
\label{bohmian-wpacket-soln}
\end{eqnarray}
with some length parameter $\ell$. It is clear that the width function $D^2$ in the wavepacket (\ref{wpacket}) vanishes if both $\hbar \rightarrow 0$ and $d\rightarrow 0$, and it is expected that in these limits the  classical solution $z=z_c(t)$ is going to be attained. This comes to mean that the parameter $\ell$ should be proportional to the width $d$, and one can set this way $\ell=d$ in (\ref{bohmian-wpacket-soln}). 

The wavepacket in (\ref{wpacket}), after shot upwards at $z=z_i$, propagates  up to the classical turning point $z=z_\cap$ and scatters back therein to fall down to $z=z_i$ in a total duration of $(\Delta t)_q^{(wp)}$. This duration is the quantum scattering time (QST) of the wavepacket. The time formula (\ref{bohmian-wpacket-soln}) leads  to the Bohmian QST 
\begin{eqnarray}
(\Delta t)_q^{(wp)}= (\Delta t)_c \left(1 + \frac{\hbar}{m \sqrt{2 g d^3}}  + {\mathcal{O}}(\hbar^2) \right)
\label{QST-wp}
\end{eqnarray}
for $\ell=d\ll z_\cap-z_i=v_i^2/2g$. In here,  $(\Delta t)_c=2v_i/g$ is the classical scattering time (CST) defined beneath the equation (\ref{z-class}). This is the average quantum scattering time. (It is ``average" in the sense that the Bohmian equation (\ref{eom-bohm}) involves probability and probability current densities and integrations over them effectively give an average duration. This becomes more evident with the Bohmian time (\ref{time-bohm}) describing the stationary-state particles.)

The quantum time formula (\ref{QST-wp}) is a proof that the equivalence principle is violated at the $\hbar$ order, where duration of penetration (tunneling) into the semi-infinite classically-forbidden region ($z>z_\cap$) is expected to be subleading since the wavepacket (\ref{wpacket}) approximates a classical particle moving on the classical 
trajectory $z_c(t)$. As a matter of fact, equivalence principle violation occurs due mainly to the dispersion of the wavepacket ($\hbar/m$ in the width function $D^2$) as was concluded also by previous studies   \cite{alis,onofrio,emelyanov}. It is clear from  the wavepacket QST in (\ref{QST-wp}) that the more QST/CST deviates from unity the stronger the violation of the equivalence principle \cite{free-fall-th1,free-fall-th2}.

\begin{table}
\caption{The QST/CST ratio. QST starts deviating from the CST at the order $\hbar$ ($\hbar^2$) for the Bohmian (Copenhagen  \cite{operator}) interpretation. Future experiments may be able to probe what interpretation of quantum behavior is allowed in nature. \label{table-0}}
\begin{flushright}
\begin{tabular}{|l|l|l|}
\hline
 {}& Bohmian interpretation & Copenhagen interpretation\\
\hline
\hline
$\frac{(\Delta t)_q^{(wp)}}{(\Delta t)_c}$ & $1+\frac{\hbar}{m \sqrt{2 g d^3}} +{\mathcal{O}}(\hbar^2) $& $1+\frac{\hbar^2}{4 m^2 d^2 v_i^2} +{\mathcal{O}}(\hbar^4) $\\ \hline
\end{tabular}
\end{flushright}
\end{table}

The deviation of the wavepacket QST from the CST  turns out to be a sensitive probe of the formulation of the quantum behavior. Table \ref{table-0} gives an example of this. Indeed, as shown by the table, for the Bohmian interpretation the deviation is an ${\mathcal{O}}(\hbar)$ effect. For the Copenhagen interpretation with the operator method \cite{operator} (similarly with the current density method \cite{alis,current}), however, the deviation is an ${\mathcal{O}}(\hbar^2)$ effect. (The formula in Table \ref{table-0} is obtained by taking $z_i\ll z_\cap$, which is not inconsistent with the Bohmian formula.) These two distinct $\hbar$-sensitivities, along with the other parametric differences, show that the future experiments may be able to probe what interpretation of quantum behavior is realized in nature. To this end, experiments  with cold atoms and neutrons  \cite{cold-neutron} may prove useful.

\section{Quantum Scattering Time: Stationary-State Particles}
In this section, we will study average scattering time of a beam of mono-energetic quantum particles from their gravitational potential, and show explicitly how this scattering duration signifies violation of the equivalence principle.  As a matter of fact, we will study a setup in which quantum particles of mass $m$ and energy $E$ are shot upwards (like a fountain) in their gravitational potential and their return time (total flight time) is recorded. As was with the wavepacket of the last section, difference between the stationary-state QST and the CST will be an indicator of the equivalence principle violation. 

It proves useful to start with the calculation of the CST. The difference from the CST in Sec. II is that this time the object of concern is a classical particle of fixed energy $E$ in the framework of the Newtonian dynamics in (\ref{newton}). This particle,  thrown upwards from $z=z_i$, rises up, turns backwards at the turning point $z=z_\cap$, and falls at $z=z_i$. This whole motion takes the total time (the CST) \cite{free-fall-th1,free-fall-th2}
\begin{eqnarray}
(\Delta t)_c &=& \int_{z_i}^{z_\cap} \frac{dz}{\sqrt{2g(z_\cap-z)}} + \int_{z_\cap}^{z_i} \frac{dz}{-\sqrt{2g(z_\cap-z)}}\nonumber\\ &=& 2 \left(\frac{2(z_\cap-z_i)}{g}\right)^{\frac{1}{2}}
    \label{time0}
    \end{eqnarray}
which is a universal duration that depends only on the gravitational acceleration $g$ and the logged height $z_\cap-z_i$.  This means that all mono-energetic classical particles approach and scatter back from their gravitational potentials in the same duration $(\Delta t)_c$  irrespective of their masses and other features. In phase space, it can be put into the form 
\begin{eqnarray}
  (\Delta t)_c = \int_{z_i}^{z_\cap} \frac{m  dz}{p_z} + \int_{z_\cap}^{z_i} \frac{m  dz}{-p_z}
    \label{time1}
\end{eqnarray}
in which $p_z=\sqrt{2m(E-V(z))}$ is momentum of the particle,  $E = m g z_\cap$ is its total energy,  and $V(z)$ is its potential energy in (\ref{pot}). This expression for $(\Delta t)_c$ expresses the march of  time in terms of the coordinate of the particle (as if a clock attached on it) \cite{challinor,abolhasani,busch,field}.

In general, the classical dynamics underlying the CST in (\ref{time1}) corresponds to  stationary-state quantum dynamics. Quantum particles obeying such dynamics possess the wavefunction 
\begin{eqnarray}
\Psi(z,t)= \psi(z) e^{-\frac{i}{\hbar}Et} 
\label{stationary}
\end{eqnarray}
whose replacement in  the time-dependent Schroedinger equation in (\ref{sch-time}) leads to
\begin{eqnarray}
-\frac{\hbar^2}{2 m} \frac{d^2 \psi(z)}{d z^2} + V(z)\psi(z) = E \psi(z) 
\label{sch}
\end{eqnarray}
as the time-independent Schroedinger equation governing  $\psi(z)$.  With the stationary-state wavefunction (\ref{stationary}),
the probability current density $J(t,z)$ in (\ref{prob-current}) takes the form 
\begin{eqnarray}
j(z)=\frac{\hbar}{2 m i} \left(\psi^\ast(z)  \frac{d}{d z}\psi(z)-\psi(z)  \frac{d}{d z}\psi^\ast(z)\right)
\label{prob-current-small}
\end{eqnarray}
and the probability density   $R(t,z)=\Psi^\ast(t,z)\Psi(t,z)$ reduces to $\rho(z)=\psi^\ast(z)\psi(z)$. Obviously, $j(z)$ must be strictly constant (though $\rho(z)$ can depend on $z$) according to the  continuity of the probability flow in (\ref{continuity}).

\begin{figure}[ht]
\includegraphics[width=8.0cm]{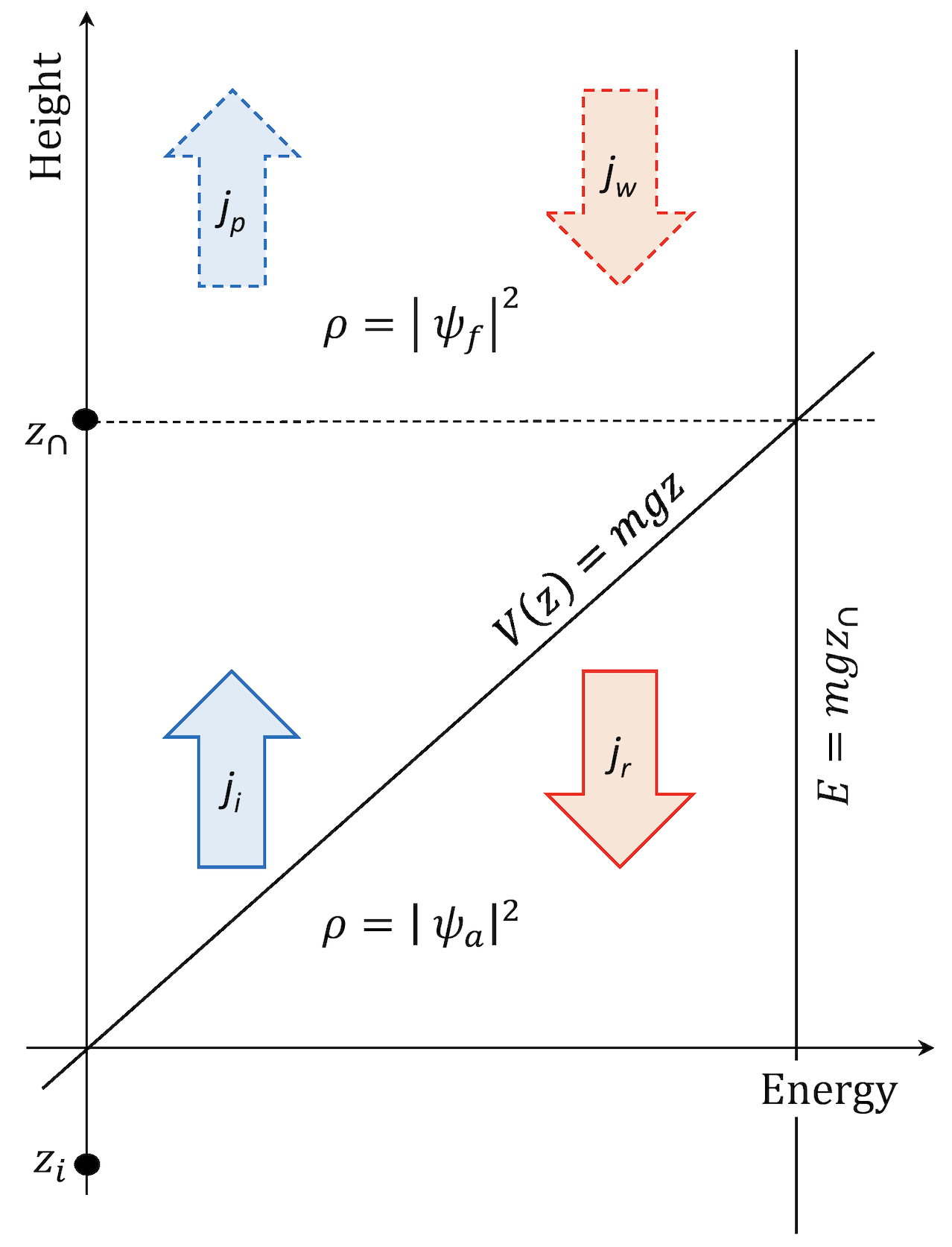}
\caption{Scattering of mono-energetic stationary-state quantum particles from their gravitational potential energy $V(z)=mgz$. In the classically-allowed region ($z<z_\cap$), the particles are distributed with the probability density $\rho=|\psi_a|^2$ and  probability undercurrents $j_i=-j_r$. Similarly, in the classically-forbidden region ($z>z_\cap$), the particles are distributed with the probability density $\rho=|\psi_f|^2$ and the probability undercurrents $j_p=-j_w$. These probability currents are obtained by judiciously splitting the wavefunctions into two complex pieces as $\psi_a(z)=\psi_i(z)+\psi_r(z)$ in the allowed region, and as $\psi_f(z)=\psi_p(z)+\psi_w(z)$ in the forbidden region. In Bohmian mechanics, quantum travel time marches with the ascribed particle position so that the total quantum scattering time is composed of the rising (up-blue full arrow), penetrating (up-blue dashed arrow), withdrawing (down-red dashed arrow), and falling (down-red full arrow) transitions.}
\label{figure-fft}
\end{figure}

The stationary-state wavefunctions like (\ref{stationary}) are tailor-made for stationary scattering events as they represent the steady flux of particles shot upwards (like a fountain) and scattered back downwards (like rain) \cite{davies}.  The problem is to define scattering time for such states in the setup depicted in Fig.~\ref{figure-fft}. To this end, as already discussed in the previous section,  Bohmian mechanics \cite{bohm,bohmian,sonego} provides a viable framework. The reason is that Bohmian mechanics assigns trajectories to quantum particles -- even to spatially spread-out stationary-state particles described by (\ref{stationary}) \cite{bohm,bohmian}. For such states, the Bohmian relation in (\ref{eom-bohm}) turns to 
\begin{eqnarray}
\frac{d z}{dt}= \frac{j}{\rho}
\label{eom-bohm-2}
\end{eqnarray}
in which $j/\rho$ is a function only of the coordinate $z$. From this one can readily construct that the travel time formula
\begin{eqnarray}
(\Delta t)_q = \int_{a}^{b} dz\,  \frac{\rho}{j} 
\label{time-bohm}
\end{eqnarray}
in which $j$ is assumed to flow from $a$ to $b$. This is the average QST corresponding to the CST in (\ref{time1}). It expresses march of the time in terms  of the probability density $\rho$ and the probability current density $j$ in the region extending from $z=a$ to $z=b$.  It is the Bohmian QST for stationary-state particles, and gives the average quantum scattering time because it is effectively the average value of the inverse  current density ($1/j$).  In applying (\ref{time-bohm}) one keeps in mind that $\rho$ can vary with $z$ but $j$ remains strictly constant.

The Schroedinger equation  (\ref{sch}) possesses the  piece-wise solution \cite{abromowitz,ryzhik}
\begin{eqnarray}
\psi(z)=\begin{cases} \psi_{a} (z)\; & {\rm for}\ z\leq z_\cap\\
\psi_{f}(z)\; & {\rm for}\ z\geq z_\cap
\end{cases}
\label{wavefunc-1}
\end{eqnarray}
in which
\begin{eqnarray}
\psi_a(z) = N \zeta^{\frac{1}{3}} 
\left(J_{\frac{1}{3}}(\zeta)+  J_{-\frac{1}{3}}(\zeta)\right)
\label{psi-allow}
\end{eqnarray}
is the wavefunction in the classically-allowed region ($z<z_\cap$), and
\begin{eqnarray}
\psi_{f}(z)=
N \zeta^{\frac{1}{3}} 
\left(I_{-\frac{1}{3}}(\zeta) - I_{\frac{1}{3}}(\zeta)\right) 
\label{psi-forbid}
\end{eqnarray}
is the wavefunction in the classically-forbidden region ($z<z_\cap$). In these solutions, $N$ is a normalization constant, and $J_{\pm 1/3}$ and  $I_{\pm 1/3}$ are the  Bessel functions of order $\pm 1/3$, with the argument
\begin{eqnarray}
\zeta= \frac{2}{3} \left(\frac{|z-z_\cap|}{L_q}\right)^{\frac{3}{2}}
\end{eqnarray}
in which 
\begin{eqnarray}
L_q=\left(\frac{\hbar^2}{2 m^2 g}\right)^{\frac{1}{3}}
\end{eqnarray}
is the natural length scale for a quantum particle under gravity. It breaks universality with $(\hbar/m)^{2/3}$ power-law. 

\subsection{Quantum Flight Time in Allowed Region}
\label{subseca}
The wavefunction $\psi_a(z)$ in (\ref{psi-allow}), describing state of the particle in the allowed region ($z<z_\cap$), can have at most a global phase. In fact, it can be taken purely real without loss of generality. It gives then  zero probability current, as follows from (\ref{prob-current-small}). This actually means that there are two equal and opposite undercurrents constituting  the stationary-state probability distribution. This implies that it must be possible to split the wavefunction $\psi_a(z)$ into two complex functions of equal and opposite currents. One can therefore write (see \cite{davies} for a similar  decomposition) 
\begin{eqnarray}
\psi_a(z)=\psi_i(z) + \psi_r(z)
\label{split}
\end{eqnarray}
in which 
\begin{eqnarray}
\psi_{i}(z)= N \zeta^{\frac{1}{3}} 
\left\{\!e^{-\frac{i\pi}{3}}J_{\frac{1}{3}}(\zeta)+ e^{\frac{i\pi}{3}} J_{-\frac{1}{3}}(\zeta)\!\right\}
\label{wavefunc-above}
\end{eqnarray}
has the positive (upward) probability current
\begin{eqnarray}
j_{i}= \frac{\hbar}{\pi m L_q} \left(\frac{3}{2}\right)^{\frac{4}{3}} \left|N\right|^2  
\label{current-i}
\end{eqnarray}
as follows from (\ref{prob-current-small}), and 
\begin{eqnarray}
\!\!\!\!\!\!\psi_{r}(z)=N \zeta^{\frac{1}{3}}\! \left\{\!\left(\!1-e^{-\frac{i\pi}{3}}\!\right)\!J_{\frac{1}{3}}(\zeta)+ \left(\!1-e^{\frac{i\pi}{3}}\!\right)\! J_{-\frac{1}{3}}(\zeta)\!\right\}
\label{wavefunc-below}
\end{eqnarray}
has the negative (downward) probability current 
\begin{eqnarray}
j_{r}=  -\frac{\hbar}{\pi m L_q} \left(\frac{3}{2}\right)^{\frac{4}{3}} \left|N\right|^2
\label{current-r}
\end{eqnarray}
as follows again from (\ref{prob-current-small}). The two currents are indeed equal in size and opposite in sign. They ensure that the wavefunction $\psi_a(z)$ is composed of an incident wave ($\psi_i$) inducing an upward probability flow and a reflected wave ($\psi_r$) creating a downward probability flow.

The decomposition of the wavefunction into two complex wavefunctions of equal-size and opposite-sign probability undercurrents has proven useful for revealing the probability underflows in the stationary-state scattering problem at hand. It worked for the allowed-region wavefunction in (\ref{psi-allow}), and it will be seen to work for the forbidden-region wavefunction (\ref{psi-forbid}) in Sec. \ref{subsecb}. It worked because the wavefunctions (\ref{psi-allow}) and (\ref{psi-forbid}) involve the Bessel functions, and Wronskians of Bessel functions lead to the required probability currents \cite{abromowitz,ryzhik,davies}. In general, decomposition becomes a necessity if  probability underflows in the stationary system are needed. The structure of the two undercurrents (equal in size and opposite in sign) determines how the decomposition should be performed but this does not guarantee uniqueness of the decomposition since there can exist different decompositions leading to the same undercurrents. Moreover, it not clear if the 
decomposition of a general wavefunction does uniquely lead to proper probability undercurrents. (This point seems to require a separate investigation. The generality and uniqueness of the decomposition is an open problem.)

Having obtained the probability currents (\ref{current-i}) and (\ref{current-r}), it is now time to compute the associated quantum flight times. It might be tempting to use the Bohmian time formula in (\ref{time-bohm}) directly. This, however, is not so easy.  The reason is that in Bohmian mechanics quantum particles are guided not by the undercurrents ($j_i$ and $j_r=-j_i$) but by the total probability current ($j_i+j_r$ which equals zero). In view of this difficulty, we introduce a Bohmian-inspired new time definition by replacing the total current in the  Bohmian time (\ref{time-bohm}) with the  $j_i$ and $j_r$ undercurrents. With this replacement, it becomes possible follow propagation of particles in the directions of the undercurrents. In this regard,  quantum particles rise from $z=z_i$ to the turning point $z =z_\cap$ within the average Bohmian-inspired time \cite{integral}
\begin{eqnarray}
  \!\!(\Delta t)_q^{(rise)} &=& \int_{z_i}^{z_\cap} \frac{\left|\psi_a(z)\right|^2}{2 j_i}  dz=  - \frac{2\pi T_q }{\left(3^{\frac{1}{3}}\Gamma\left(\frac{1}{3}\right)\right)^2} \nonumber\\
  &+& 2\pi T_q \left(\beta_q\left({\rm Ai}(-\beta_q)\right)^2 + \left({\rm Ai}^\prime(-\beta_q)\right)^2 \right)
  \label{time-q-rise}
\end{eqnarray} 
in which $\beta_q$ is quadratic in $(\Delta t)_c$
\begin{eqnarray}
\beta_q = \left(\frac{(\Delta t)_c}{4 T_q}\right)^2
\end{eqnarray}
and involves 
\begin{eqnarray}
T_q=\left(\frac{\hbar}{4 m g^2}\right)^{\frac{1}{3}}
\label{q-time-scale}
\end{eqnarray}
as the natural time scale for a quantum particle under gravity. In the rise time (\ref{time-q-rise}), at the right-hand side, the function ${\rm Ai}(\dots)$ is the Airy function of the first kind, and ${\rm Ai}^\prime(\dots)$ is its derivative \cite{abromowitz,ryzhik}. 

In parallel with the rise time above, quantum particles are found to fall from the turning point $z_\cap$ to $z_i$ within the average Bohmian-inspired time \cite{integral}
\begin{eqnarray}
 (\Delta t)_q^{(fall)} &=& \int_{z_\cap}^{z_i} \frac{\left|\psi_a(z)\right|^2}{2j_r}  dz = (\Delta t)_q^{(rise)} 
  \label{time-q-fall}
\end{eqnarray}
where the  $1/2$ factor in the integrands of $(\Delta t)_q^{(rise)}$ and $(\Delta t)_q^{(fall)}$ is there to avoid double counting while keeping the interference terms between $\psi_i$ and $\psi_r$. These two times give the average flow durations in the  classically-allowed region in Fig.~\ref{figure-fft}.

\subsection{Quantum Flight Time in Forbidden Region}
\label{subsecb}
The wavefunction $\psi_f(z)$ in (\ref{psi-forbid}), describing the state of the particle in the classically-forbidden region ($z>z_\cap$), can have at most a global phase. It can in fact be taken real (like $\psi_a(z)$ in the classically-allowed region) without loss of generality. It possesses zero probability current as follows from (\ref{prob-current-small}). As in Sec. \ref{subseca}, this zero current can be structured as being composed of two equal and opposite undercurrents by an appropriate splitting of the stationary-state wavefunction into two complex wavefunctions. One can  write therefore
\begin{eqnarray}
\psi_f(z)=\psi_p(z) + \psi_w(z)
\label{split-f}
\end{eqnarray}
in parallel with (\ref{split}) such that 
\begin{eqnarray}
\psi_{p}(z)= N i \zeta^{\frac{1}{3}} 
\left\{\!e^{\frac{i\pi}{6}}I_{\frac{1}{3}}(\zeta)+ e^{-\frac{i\pi}{6}} I_{-\frac{1}{3}}(\zeta)\!\right\}
\label{wavefunc-penetrate}
\end{eqnarray}
has the positive (upward) probability current
\begin{eqnarray}
j_{p}= \frac{\hbar}{\pi m L_q} \left(\frac{3}{2}\right)^{\frac{4}{3}} \left|N\right|^2  
\label{current-p}
\end{eqnarray}
as follows from (\ref{prob-current-small}), and 
\begin{eqnarray}
\!\!\!\!\!\!\!\!\psi_{w}(z)=-N \zeta^{\frac{1}{3}}\! \left\{\!\left(\!1-e^{-\frac{i\pi}{3}}\!\right)\!I_{\frac{1}{3}}(\zeta)- \left(\!1-e^{\frac{i\pi}{3}}\!\right)\! I_{-\frac{1}{3}}(\zeta)\!\right\}
\label{wavefunc-withdraw}
\end{eqnarray}
has the negative (downward) probability current 
\begin{eqnarray}
j_{w}=  -\frac{\hbar}{\pi m L_q} \left(\frac{3}{2}\right)^{\frac{4}{3}} \left|N\right|^2
\label{current-w}
\end{eqnarray}
as follows again from (\ref{prob-current-small}). The two undercurrents are indeed equal in size and opposite in sign.  They ensure thus that the wavefunction $\psi_f(z)$ is composed of a penetrating evanescent wave ($\psi_p$, decaying towards $z=\infty$) inducing an upward probability flow and a withdrawing evanescent wave ($\psi_w$, decaying towards $z=z_\cap$) creating a downward probability flow.

Having derived the probability currents (\ref{current-p}) and (\ref{current-w}), average quantum flight times in the classically-forbidden region ($z>z_\cap$) can now be computed by using the Bohmian-inspired time formula in Sec. \ref{subseca}. Indeed, it turns out that an evanescencing quantum particle penetrates into the forbidden region for an average Bohmian-inspired time \cite{integral}
\begin{eqnarray}
  \!\!(\Delta t)_q^{(penetrate)} &=& \int_{z_\cap}^{\infty} \frac{\left|\psi_f(z)\right|^2}{2j_p}  dz=  
  \frac{2\pi T_q}{\left(3^{\frac{1}{3}}\Gamma\left(\frac{1}{3}\right)\right)^2}
  \label{time-q-penetrate}
\end{eqnarray} 
whose right-hand side is set by ${\rm Ai}^\prime(0)$ \cite{abromowitz,ryzhik}. 

In parallel with the penetration time above, quantum particles withdraw back to the turning point $z_\cap$ in the average Bohmian-inspired time \cite{integral}
\begin{eqnarray}
\!\!(\Delta t)_q^{(withdraw)}&=& \int_{\infty}^{z_\cap} \frac{\left|\psi_f(z)\right|^2}{2j_w}  dz=  
  \frac{2\pi T_q}{\left(3^{\frac{1}{3}}\Gamma\left(\frac{1}{3}\right)\right)^2}
  \label{time-q-withdraw}
\end{eqnarray}
where the factor $1/2$ in the integrands of $(\Delta t)_q^{(penetrate)}$ and $(\Delta t)_q^{(withdraw)}$ is placed to prevent double counting while keeping the cross terms between $\psi_p$ and $\psi_w$. These two times sum up to the total time spent in  the classically-forbidden region (as depicted in Fig.~\ref{figure-fft}).

Before going any further, it proves instructive to discuss  time spent in the classically-forbidden region ($z>z_\cap$) also in the dwell time formulation \cite{dwell-time1}. In this formulation, quantum particles of incidence current $j_{inc}$ spend a time  \cite{dwell-time1,dwell-time2}
\begin{eqnarray}
(\Delta t)^{(dwell)} = \frac{1}{j_{inc}} \int_a^b dz\, \rho_f
\label{dwell}
\end{eqnarray} 
in a classically-forbidden region extending from $z=a$ to $z=b$, with the probability density $\rho_f$. This time formula differs from the Bohmian time (\ref{time-bohm}) by the fact that the current $j_{inc}$ is the incident current, not the current in the forbidden region extending from $a$ to $b$. Despite this, explicit calculation shows that the dwell time  satisfies the relation  \cite{integral}
\begin{eqnarray}
  (\Delta t)_q^{(dwell)} 
  = (\Delta t)_q^{(penetrate)} + (\Delta t)_q^{(withdraw)}
  \label{time-q-dwell}
\end{eqnarray}
after letting $\rho^{f} \rightarrow \left|\psi_f(z)\right|^2$ and
$j_{inc}\rightarrow j_i$ in (\ref{dwell}), where  
$\psi_f(z)$ and $j_i$ are defined in (\ref{split-f}) and  (\ref{current-i}), respectively. The relation (\ref{time-q-dwell}) gives an independent confirmation of the Bohmian-inspired travel time formula (splitting of the wavefunction in two complex pieces as in (\ref{split-f}) and use of the respective currents (\ref{current-p}) and (\ref{current-w})). 

\subsection{Quantum Scattering Time}
On physical grounds,  QST  is fundamentally different than the CST in (\ref{time1}). Indeed, while quantum particles perform a ``rise-penetrate-withdraw-fall" motion the classical particles perform a  simple ``rise-turn-fall" motion. The reason is that there is essentially no turning point for a quantum particle as it is always able to penetrate  into  the $z >z_\cap$ domain (as in (\ref{wavefunc-penetrate})) and withdraw back (as in (\ref{wavefunc-withdraw})) as a semi-infinite tunneling transition induced by evanescent waves. (This effect is expected to be subleading for a wavepacket as discussed in Sec. \ref{secwaveacket}.) All this implies that the QST is composed of four segments
\begin{eqnarray}
(\Delta t)_q &=& (\Delta t)_q^{(rise)} + (\Delta t)_q^{(penetrate)}\nonumber\\ &+& (\Delta t)_q^{(withdraw)} +(\Delta t)_q^{(fall)}
\label{tofff}
\end{eqnarray}
as an ordered set of transitions depicted in Fig.~\ref{figure-fft}. Now, collecting the individual time intervals from (\ref{time-q-rise}),  (\ref{time-q-penetrate}), (\ref{time-q-withdraw}) and (\ref{time-q-fall}), this $(\Delta t)_q$ formula leads to the QST/CST ratio
\begin{eqnarray}
\frac{(\Delta t)_q}{(\Delta t)_c} &=& 
 \pi \sqrt{\beta_q}\left(\! {\rm Ai}(-\beta_q)\right)^2 + \frac{\pi}{\sqrt{\beta_q}}\left({\rm Ai}^\prime(-\beta_q)\right)^2
\label{tofff2}
\end{eqnarray}
as because $(\Delta t)_q^{(penetrate)} + (\Delta t)_q^{(withdraw)}$ cancels out  the constant part in $(\Delta t)_q^{(rise)} + (\Delta t)_q^{(fall)}$. This exact result shows how QST differs from the CST as a function of the universality-breaking parameter $\hbar/m$. This dependence on $\hbar/m$ ensures that QST/CST is an unambiguous vestige of equivalence principle violation. More specifically, the more QST/CST deviates from unity the stronger the violation of the equivalence principle \cite{free-fall-th1,free-fall-th2}. 

\begin{table}
\caption{The quantum characteristic time scale $T_q$ in (\ref{q-time-scale}) and quantum mechanical collision time in  (\ref{no-rise-no-fall}) for the electron and neutron. In general, $T_q({\rm atom}) \approx T_q({\rm neutron})\times A^{-1/3}$ for an atom with mass number $A$. \label{table-1}}
\begin{flushright}
\begin{tabular}{|l|l|l|l|}
\hline
Particle & Mass (kg) & $T_q$ (s)& $(\Delta t)_q[z_i=z_{\cap}]$(s)\\
\hline
\hline
Electron & $9.109\times 10^{-31}$& $1.496\times 10^{-8}$ & $1.259\times 10^{-8}$\\ \hline
Neutron & $1.674\times 10^{-27}$& $1.221\times 10^{-9}$& $1.028\times 10^{-9}$\\ \hline
\end{tabular}
\end{flushright}
\end{table}

\begin{figure}[ht]
\includegraphics[width=8.6cm]{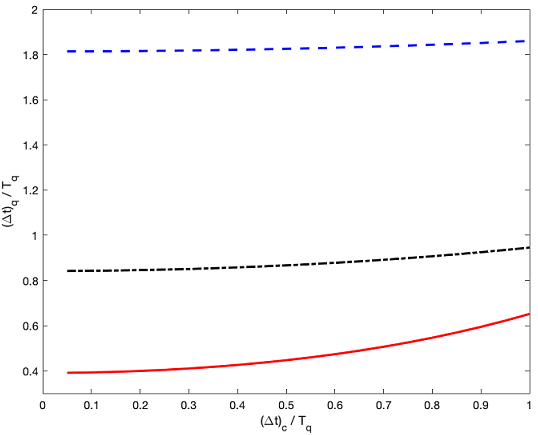}
\caption{Variation of the QST ($(\Delta t)_q/T_q$)  with the CST  ($(\Delta t)_c/T_q$)  for quantum particles having masses $m$ (dot-dashed black), $10\times m$ (full red) and $m/10$ (dashed blue). It is clear that, in each case, QST remains nonzero even when the CST vanishes exactly. These QST values at $(\Delta t)_c=0$ (namely, $z_i=z_\cap$) are a proof that the quantum particles possess a finite collision time at the turning point, which serves as an indicator of the equivalence principle violation.}
\label{figure-nonzeroTOFFF}
\end{figure}

\begin{figure}[ht]
\includegraphics[width=8.7cm]{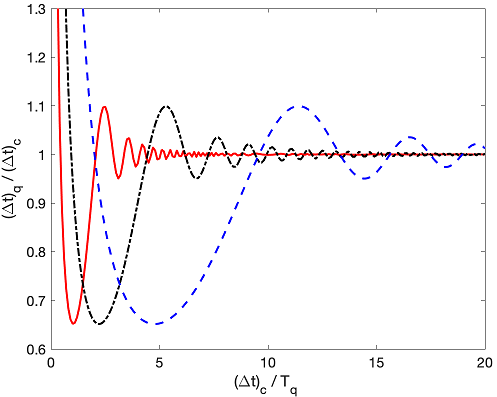}
\caption{Variation of the QST/CST ($(\Delta t)_q/(\Delta t)_c$) with the CST ($(\Delta t)_c/T_q$) for quantum particles having masses $m$ (dot-dashed black), $10\times m$ (full red) and $m/10$ (blue dashed). It is clear that each QST relaxes to the CST in an oscillatory fashion such that lighter (heavier) the particle slower (faster) the relaxation. Equivalence principle violation is pronounced at low values of $(\Delta t)_c/T_q$.}
\label{figure-periodic}
\end{figure}

One physically important regime of QST in (\ref{tofff2}) is the short-flight regime namely $z_i \rightarrow z_\cap$ limit. In this regime, the particle starts already  at the turning point $z=z_{\cap}$, penetrates into the semi-infinite barrier for a duration $(\Delta t)_q^{(penetrate)}$, and reappears at the turning point after a time lapse of  $(\Delta t)_q^{(withdraw)}$. In this short-flight limit, the CST vanishes identically as follows from (\ref{time0}) ($(\Delta t)_c[z_i - z_{\cap}] = 0$) but the QST takes the nonzero value
    \begin{eqnarray}
(\Delta t)_q[z_i = z_{\cap}] &=& (\Delta t)_q^{(penetrate)} + (\Delta t)_q^{(withdraw)}\nonumber\\
&=&\frac{4\pi T_q}{\left(3^{\frac{1}{3}}\Gamma\left(\frac{1}{3}\right)\right)^2}
\label{no-rise-no-fall}
\end{eqnarray}
which shows that the quantum particle wanders in the classically-forbidden region for a finite duration. This wandering is due to particle's penetration into and withdrawal from  the $z> z_\cap$ domain. It  turns out that  the tunneling into the semi-infinite potential barrier $V(z)>E$ makes quantum particle to acquire a finite collision duration at the turning point.  Indeed, as depicted in Fig.~\ref{figure-nonzeroTOFFF} for particles of  masses $m$ (dot-dashed black), $10\times m$ (full red) and $m/10$ (dashed blue), QST remains nonzero even when the CST vanishes (zero-flight limit). This means that each particle spends a finite time at the turning point corresponding to collision duration with the gravitational potential barrier.  By definition, $(\Delta t)_q[z_\cap=z_i]$ is an ${\mathcal{O}}\left((\hbar/m)^{1/3}\right)$ quantum effect, and varies from particle to particle as exemplified in Table \ref{table-1} for the electron and the neutron.   

Another physically important regime of QST in (\ref{tofff2}) is the high-flight regime namely $z_\cap-z_i \gg L_q$ regime. In this limit, on physical grounds, one expects  QST to approach to the CST. Indeed, for $z_\cap-z_i \gg L_q$ the exact QST/CST in (\ref{tofff2}) takes the form
\begin{eqnarray}
\frac{(\Delta t)_q[z_\cap-z_i \gg L_q]}{(\Delta t)_c} &=&   1-  \frac{\cos\alpha_q}{3 \alpha_q} + {\mathcal{O}}\left((\hbar/m)^2\right)
\label{tofff3}
\end{eqnarray}
where $\alpha_q = \frac{4}{3}  \left(\beta_q\right)^{\frac{3}{2}}$ 
parametrizes the universality-breaking quantum contributions, which vary from particle to particle via  $\alpha_q \propto m/\hbar$. The parameter $\alpha_q$ gives information about equivalence principle violation by a measurement of QST/CST for long flights.  In general, heavier the particle smaller the quantum contribution as revealed by the $T_q$ values in Table \ref{table-1}. Direct calculation reveals that the high-flight QST in (\ref{tofff3}) holds for distances grater than $0.274\ {\rm fm}$ ($0.183\ {\rm fm}$) for electrons (neutrons).

Depicted in Fig.~\ref{figure-periodic} is QST as a function of the CST for particles of  masses $m$ (dot-dashed black), $10\times m$ (full red) and $m/10$ (dashed blue). The plot extends from short-flight to high-flight regime as $(\Delta t)_c/T_q$ increases. It is clear that the QST exhibits strong swings at low $(\Delta t)_c/T_q$, which can be detected experimentally by using beams of different energies. It is also clear that the QST relaxes to the CST at large $(\Delta t)_c/T_q$ in an oscillatory fashion such that lighter (heavier) the particle slower (faster) the relaxation. Evidently, equivalence principle violation becomes stronger at low $(\Delta t)_c/T_q$. This reduction of the QST to the CST at large $(\Delta t)_c/T_q$ is also what is emphasized  in \cite{davies} by Davies. The operator approach in \cite{operator} is valid only if the particle does not reach $z_\cap$ and is thus not possible to contrast with the results here. Nevertheless, both \cite{davies} and \cite{operator} find ${\mathcal{O}}(\hbar)$ and higher-order (positive or negative) corrections to the CST. 

\section{Experimental Determination}
Universality of free-fall  has been under experimental exploration for decades \cite{test1,test2,test3}. In the last decade experiments have  diversified and reached higher precision levels \cite{rubidium1,rubidium2,rubidium3,rubidium4,rubidium5,rubidium6}. The experiments with cold atoms are particularly promising. It is likely that  experiments as such, including cold neutrons \cite{cold-neutron}, can start measuring flight times  of quantum particles in near future. Such scattering experiments can be conducted reliably under ultra-high vacuum conditions corresponding to pressures about $10^{-10}\ {\rm Pa}$ and mean free paths about $10^5\ {\rm m}$. 

The present work reports actually two classes of new results. The first concerns scattering time of wavepackets \cite{rubidium1,rubidium2,rubidium3,rubidium4,rubidium5,rubidium6}. It was analyzed in Sec. \ref{secwaveacket}, with the main result that a proper measurement of the scattering time can distinguish between the Bohmian and Copenhagen interpretations of the quantum behavior. Indeed,   equivalence principle violation is of size $\hbar$ ($\hbar^2$) in the Bohmian (Copenhagen) approach, wavepacket QST becomes a new distinguishing quantity after the quantum backflow \cite{backflow1,backflow2,backflow3}. Future experiments might probe what interpretation is realized in nature. 

The second class of new results concern scattering times of mono-energetic beam of stationary-state particles. One can shoot such  particles upwards, make them scatter off from their gravitational potential, and measure their average return times. The  determinations of $(\Delta t)_q[z_\cap=z_i]$ and $(\Delta t)_q[z_\cap-z_i \gg L_q]$ are particularly important for various reasons. Indeed, experimental verification of $(\Delta t)_q[z_\cap=z_i]$ in (\ref{no-rise-no-fall}) would ensure that
\begin{enumerate}
\item  quantum free-fall is not universal, 
\item quantum tunneling takes finite time, and 
\item quantum travel time could be Bohmian.
\end{enumerate}
Experimental confirmation of $(\Delta t)_q[z_\cap-z_i \gg L_q]$, on the other hand, would ensure that
\begin{enumerate}
\item quantum free-fall is not universal, 
\item quantum particles can fall much faster or slower than the classical particles, and 
\item universal classical free-fall times are attained for long flights. 
\end{enumerate}
In general, QST is around  
nanoseconds for cold neutrons and significantly shorter for cold atoms (cesium, potassium, rubidium, and the like). Table \ref{table-1},  Fig. \ref{figure-periodic} and Fig. \ref{figure-nonzeroTOFFF} provide the necessary information. These scattering times should give an idea about the precision goal in future experiments. 

\section{Conclusion}
In this work, we have performed a systematic study of the scattering times of quantum particles from their gravitational potentials. We have utilized the opportune Bohmian mechanics as it ascribes trajectories to quantum particles. We have first analyzed scattering times of wavepackets in the Bohmian formalism in a way involving the equivalence principle violating ratio $\hbar/m$.  We have found that scattering times can distinguish between the Bohmian and Copenhagen interpretations. 

We have next analyzed mono-energetic stationary-state particles corresponding to steady flux of quantum particles, and  shown that their quantum and classical scattering times differ from each other in a way involving the equivalence principle violating ratio $\hbar/m$. We have analyzed the quantum scattering time in short- and high-flight regimes and low- and high-mass limits, and found explicit expressions testable by appropriate scattering experiments. It turns out that experiments with different particle energies and and different particle masses seem to have good potential to test the quantum violation of the equivalence principle.  The formula found can prove useful for both theoretical and experimental tests of the equivalence principle in quantum systems. 

Experimental determination of the quantum scattering time of wavepackets can determine what interpretation of the quantum behavior is realized in nature. The scattering times of stationary-state particles, on the other hand, can put an end to the quest for the correct formula for traversal and tunneling times in quantum theory. And analyses of the tunnel ionization of atoms can provide a cross check for experimental data \cite{phase-1,tugrulla,yakaboylu,biz}. Fundamentally, quantum scattering time, if measured accurately, can innovate our conception of time in quantum theory, with widespread implications for tunneling-enabled processes.

\begin{acknowledgments}
This work is supported by the IPS Project B.A.CF-20-02239 at Sabanc{\i} University. The author is grateful to conscientious reviewers for their useful criticisms, questions and suggestions. 
\end{acknowledgments}


\end{document}